\documentclass{spie}
\usepackage{graphicx}
\usepackage{amssymb}

\title{Observing faint targets with MIDI at the VLTI\\
The MIDI AGN Large Programme\footnote{\quad Based on observations at the European Southern Observatory VLTI, Large Programme ID 184.B-0832} experience}

\author{Leonard Burtscher\supit{a}\supit{b}, Konrad R. W. Tristram\supit{c}, Walter J. Jaffe\supit{d}, Klaus Meisenheimer\supit{b}
\skiplinehalf
\supit{a}	burtscher@mpe.mpg.de, Max-Planck-Institut f\"ur extraterrestrische Physik,
		Postfach 1312, Gie\ss enbachstr., 85741 Garching, Germany\\
\supit{b}	Max-Planck-Institut f\"ur Astronomie,
		K\"onigstuhl 17, 69117 Heidelberg, Germany\\
\supit{c}	Max-Planck-Institut f\"ur Radioastronomie,
		Auf dem H\"ugel 69, 53121 Bonn, Germany\\
\supit{d}	Sterrewacht Leiden, Universiteit Leiden,
		Niels-Bohr-Weg 2, 2300 CA Leiden, The Netherlands
}

\newcommand{\jnlref}[1]{{ \textsf #1}}
\newcommand\araa{\jnlref{ARA\&A}}%
\newcommand\apj{\jnlref{ApJ}}%
\newcommand\apjl{\jnlref{ApJ}}%
\newcommand\aap{\jnlref{A\&A}}%
\newcommand\mnras{\jnlref{MNRAS}}%
\newcommand\nat{\jnlref{Nature}}%
\newcommand{\um}{\ensuremath{\mu\mathrm{m}}\,}
\newcommand{\arcsec}{''}

\begin{document}
\maketitle

\begin{abstract}
In order to put MIDI/VLTI observations of AGNs on a significant statistical basis, the number of objects had to be increased dramatically from the few prominent bright cases to over 20. For this, correlated fluxes as faint as $\approx$ 150 mJy need to be observed, calibrated and their errors be estimated reliably. We have developed new data reduction methods for the coherent estimation of correlated fluxes with the {\em Expert Work Station} (EWS). They increase the signal/noise of the reduced correlated fluxes by decreasing the jitter in the group delay estimation. While correlation losses cannot be fully avoided for the weakest objects even with the improved routines, we have developed a method to simulate observations of weak targets and can now detect --- and correct for --- such losses.
We have analyzed all sources of error that are relevant for the observations of weak targets. Apart from the photon-noise error, that is usually quoted, there is an additional error from the uncertainty in the calibration (i.e.\ the conversion factor).
With the improved data reduction, calibration and error estimation, we can consistently and reproducibly observe fluxes as weak as $\approx$ 150 mJy with an uncertainty of $\approx$ 15 \% under average conditions.
\end{abstract}


\section{Introduction: Scientific motivation}

There is little doubt that the energy-releasing central engine of an Active Galactic Nucleus (AGN) consists of a black hole surrounded by an accretion disk that produces copious amounts of radiation from infrared to X-Rays. At a distance of several milli parsec, gas clouds get irradiated by this light and show Doppler-broadened broad emission lines that are observed in the direct light in so called ``type 1'' AGNs. In others, called ``type 2'', only narrow emission lines are observed in the direct light but the broad lines are weakly detectable in polarized, i.e.\ scattered, light. Unified schemes of AGNs \cite{antonucci1993} ascribe this difference in the optical spectra of AGNs to different lines of sight towards the symmetry axis of the system and require a toroidal dusty structure -- the so called ``torus'' -- that intercepts the radiation from the accretion disk and reprocesses it in the infrared.

In observations with a single 8m telescope (``single-dish observations''), the torus is unresolved. While single-dish spectra\cite{hoenig2010, ramosalmeida2011} are able to constrain a number of parameters for models of clumpy tori\cite{nenkova2002,hoenig2006,nenkova2008,schartmann2008,hoenig2010}, it is challenging to isolate the torus emission from the emission of a nuclear starburst and from synchrotron radiation of a nuclear jet in unresolved observations. In order to resolve the torus emission, spatial resolutions of better than 100 milli arcseconds (mas) are needed. The MID-infrared Interferometric instrument (MIDI) at the Very Large Telescope Interferometer (VLTI) on Cerro Paranal, Chile, is worldwide unique to observe at such high resolutions in the atmospheric $N$ band ($\lambda \approx$ 8 -- 13 $\um\!$) where the torus SED peaks.

The MIDI AGN programme has found parsec-scale dust structures that can be associated with the ``torus'' from the unified model \cite{jaffe2004,tristram2007b,raban2009,burtscher2009,tristram2012}. A study using GTO time \cite{tristram2009} showed that objects fainter than the two brightest targets can still be observed and resolved with MIDI and a recent study of six type 1 objects \cite{kishimoto2011b} hinted at more compact dust structures for higher AGN luminosities. However, only few sources have been studied initially and the implications on the prevalent astrophysical questions -- e.g.\ are ``type 1'' and ``type 2'' tori different? Is there a connection between torus properties and AGN activity? How does the torus size vary with AGN luminosity? -- were limited by the low number of sources. It was clear that a large and systematic observational campaign was needed to collect the basic observational information necessary to understand dusty tori on a statistical basis. This is the aim of the MIDI AGN Large Programme (LP). It was also clear, however, that the sample size can only be increased by observing targets that are fainter than ESO's official limiting magnitude for visibility observations with MIDI on the UTs ($N=4^{\rm mag}$ or $F_{\nu}$ = 1 Jy for an unresolved object).

The data and first scientific results of the MIDI AGN LP will be presented elsewhere (L. Burtscher et al., in preparation). The topics of this proceedings article are the observing strategy and data reduction method necessary to successfully observe and calibrate faint (100 mJy $< F_{\nu} <$ 1 Jy for an unresolved source) objects with MIDI.

This paper is structured as follows: First we describe what an observer can do at the telescope to ensure that faint targets can be optimally calibrated. Next, an outline of the data reduction method is given and the modifications and specific problems for weak targets are described. Lastly, we discuss an improved error estimation for weak sources and summarize.

\section{Observing method}
\label{sec:obs}

   \begin{figure}
   \begin{center}
   \begin{tabular}{c}
   \includegraphics[height=4cm]{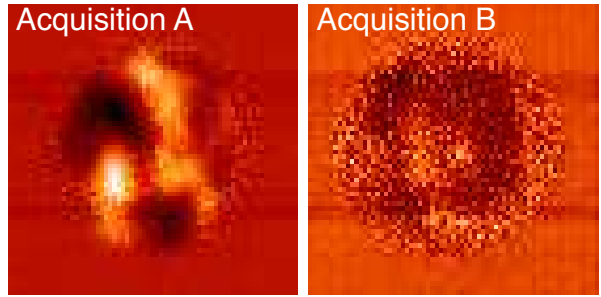}
   \end{tabular}
   \end{center}
\caption{ \label{fig:acq} Typical background in chopped acquisition images of a weak target. These two images were taken nearly simultaneously in the night of 2010-01-31 at 4:45 UTC on NGC 4507 with 8000 exposures of 4 ms. The target is a type 2 AGN with $F_{\nu} (12 \um) \approx$ 0.6 Jy. Not only is the background still very high and the target only barely visible near the center of beam B, but the background in beam A is also very inhomogeneous, thereby completely masking the target.}
   \end{figure} 

A full observing sequence with MIDI consists of the following steps for both the calibrator and the science target:
\begin{enumerate}
	\item {\bf Acquisition images} can be taken upon the observer's request in order to verify that the beam overlap is good (see Fig.~\ref{fig:acq}).
	\item The {\bf fringe search} is used to locate the position of zero optical path difference (ZOPD) based on an initial guess using a baseline model.
	\item During {\bf fringe track}, the fringe pattern is scanned with the MIDI-internal piezo motors and the fringe signal is recorded.
	\item Lastly, two {\bf photometry} observations (= single-dish spectra; one per telescope) can be taken to determine the total flux (density) of the target.
\end{enumerate}

As one can see in Fig.~\ref{fig:acq}, one cannot hope to constrain the overlap or judge other observing conditions using acquisition images for faint targets. The single-dish spectra (``MIDI photometries'') are also very hard to calibrate since they suffer from the same strong and variable background that can usually not be completely removed. While the origin of these fluctuations is still unclear, they most likely arise {\em inside} the VLTI optical train and may arise from emission from the telescope structure reflected into the optical path, imperfect corrections by the AO-system MACAO or turbulence in the VLTI tunnels. Many such observations have to be performed to obtain a good estimate of the spectrum of the source. The uncertainty in the single-dish spectra is probably best described as a composite between photon noise and background fluctuations, where the latter are dominant for the weakest observable sources\cite{burtscher2011}. Since the uncertainty in MIDI photometry is so large for weak sources, it is advisable to skip this step in the observing sequence alltogether. This changes the way, the data have to be calibrated.

The default calibration for MIDI is, for each wavelength bin,

\[V = \frac{V_{\rm target}}{V_{\rm ins}} = \frac{C_{t,0} \cdot T}{P_{t,0} \cdot T} \cdot \frac{P_{c,0} \cdot T'}{C_{c,0} \cdot T'} = \frac{C_t/P_t}{C_c/P_c}.\]

where $V$ is the calibrated visibility of the target and $V_{\rm target}$, $V_{\rm ins}$ are the raw visibility of the target and the instrumental visibility as measured from the (unresolved) calibrator, respectively. $C$ and $P$ stand for the correlated flux and photometry counts of the \underline{{\em t}}arget and \underline{{\em c}}alibrator, respectively. The index 0 stands for the (hypothetical) ``true'' count rate without atmospheric and VLTI absorptions and $T$ and $T'$ are the transmission functions at two instants in time. It is assumed that the same $T$ applies for both the correlated flux and the photometry observations, since they are taken very close in time, i.e.\ both $T$ and $T'$ cancel out. Instrumental effects, that change much slower than the transfer function, are removed by observing a calibrator at a later time. However, for weak targets, this calibration method is problematic since the uncertainty in the photometry is very large.

An alternative calibration method (``direct flux calibration'') exists,

\[F_{\nu} = \frac{C_{t,0} \cdot T}{C_{c,0} \cdot T'} \cdot F_{\nu, c} \cdot V_{\rm cal},\]

where $F_{\nu}$ is the correlated flux of the target, $F_{\nu,c}$ is the flux of the calibrator and $V_{\rm cal}$ the visibility of the calibrator (typically very close to unity in the $N$ band). With this calibration method, the noisy total flux measurement of the target, $P_t$, is not required anymore. On the other hand, since the division involves two quantities taken many minutes apart (the correlated flux of the target and of the calibrator), the transfer functions can no longer be assumed to be identical and one needs to estimate the fraction $T/T'$.

Therefore, the observing sequence should be optimized to switch as fast as possible between target and calibrator fringe track. This can be achieved by concentrating on fringe track observations and omitting the single-dish observations (acquisition, photometry). Fringe searches should only be taken if the baseline model is not sufficient for MIDI to find the fringe during the first few seconds of the track\footnote{The fringe search may be omitted for both the calibrator and the target if the baseline model is excellent, but can in most cases be omitted for one of the two. Usually calibrator and science target are close on sky so that the differential deviations from the baseline model are within the coherence envelope. The ZOPD position may be found by the MIDI-internal fringe tracker in PRISM mode (coherence length $\Lambda_{\rm coh} = R \cdot \lambda \approx 300 \um$) even when starting the track as far away from ZOPD as 4 $\Lambda_{\rm coh}$ for bright sources (calibrators). If the fringe search is omitted, it is advisable to set the OPD position where the fringe track starts to a lower OPD value than the previously found value of ZOPD. This way, MIDI will ``run into the fringe'', since the MIDI-internal fringe tracker starts increasing the OPD if the fringe is not found.}.

If only fringe tracks are taken, the time for a calibrated $(u,v)$ point can be reduced to approximately 16 minutes, i.e.\ the time for slewing, acquisition and integration can be reduced to approximately 8 minutes per target, under average conditions (see Fig. \ref{fig:gain}). Thus, a sequence of CAL--SCI--CAL--SCI--CAL, i.e.\ two science target fringe tracks embedded in calibrator tracks, takes about the same time as the currently given time for a calibrated $(u,v)$ point (50 min)\cite{dewit2012}, but is a much better estimate for the correlated flux of a weak target than the default observing procedure. The optimized observing procedure, involving only correlated flux, has recently been implemented by ESO as ``correlated flux mode''\cite{dewit2012}.

In order to calculate a visibility from $F_{\nu}$, a single-dish spectrum, e.g.\ from VISIR/VLT, is required in addition to the MIDI observations.

\section{Data reduction for weak targets}

Weak source observations with MIDI are best reduced using the method of coherent integration\cite{meisner2004} as it provides a higher signal/noise compared with the technique of power spectrum estimation for each frame. The former method has the additional advantage that any noise terms average to complex zero, whereas in the $V^2$ case, a constant offset remains that is hard to determine for weak sources. The {\em Expert Work Station} (EWS)\cite{jaffe2004b}, currently in version 2.0 (beta)\footnote{For this article, the EWS ``snapshot'' version of 25 Jan 2012 was used.}, implements the method of coherent integration for MIDI data. The software is free under an open source license and its source can be downloaded from \href{http://home.strw.leidenuniv.nl/~jaffe/ews/index.html}{http://home.strw.leidenuniv.nl/$\sim$jaffe/ews/index.html} where a technical description of the routines is also given.

The data reduction for MIDI data with EWS consists of the reduction of the fringe track data and the reduction of the single-dish ``photometry'' data (if available). The procedure is then repeated for the calibrator. The photometric data reduction has been described in detail\cite{tristram2007} and there is not much one can do to improve the MIDI photometry for weak sources. The determination of the correlated flux on the other hand has changed significantly since the original description\cite{burtscher2011}; here the new features of EWS 2.0 are highlighted.

\subsection{Outline}

\begin{figure}
\begin{center}
\begin{tabular}{c}
\includegraphics[width=8cm, angle=270]{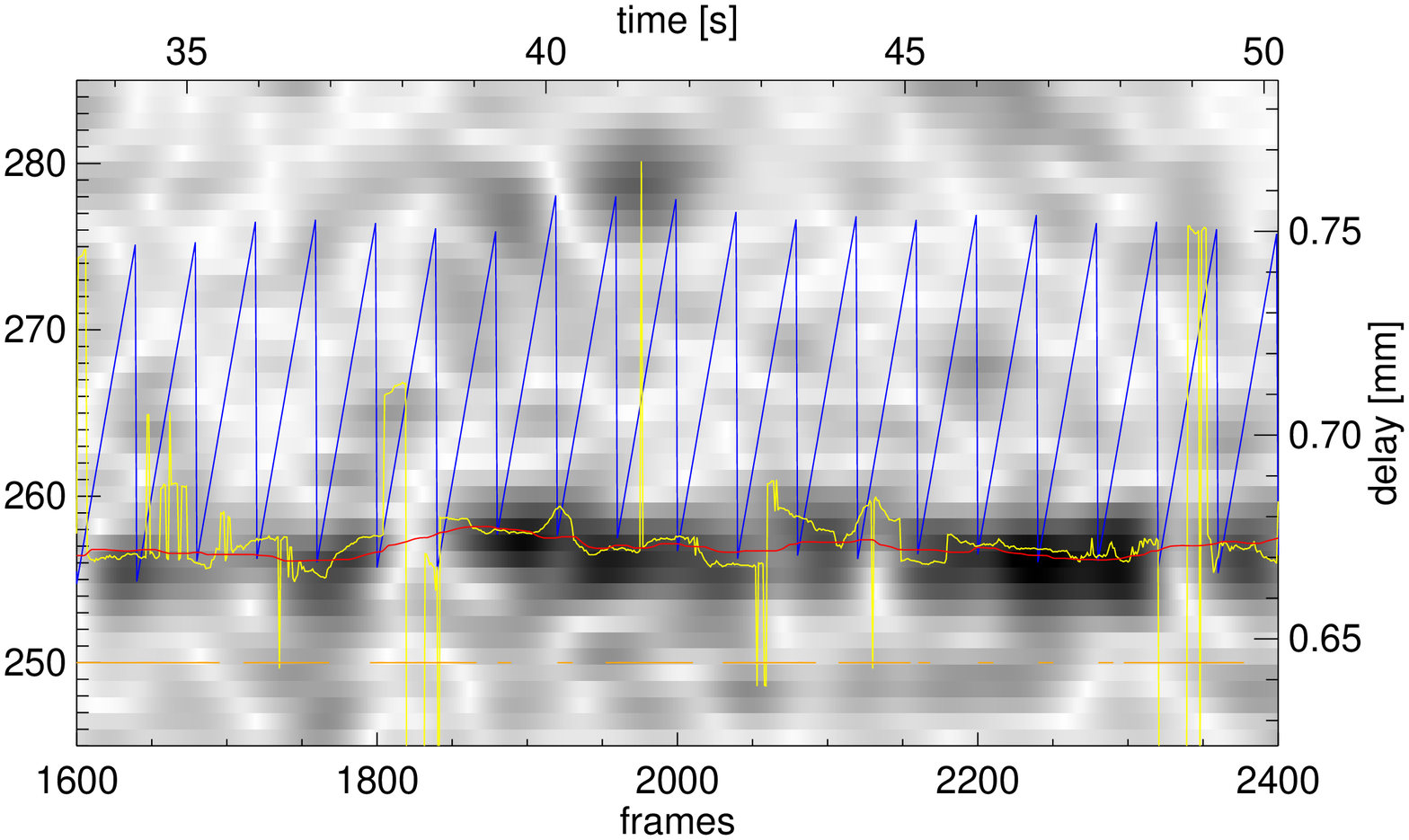}\\
\includegraphics[width=8cm, angle=270]{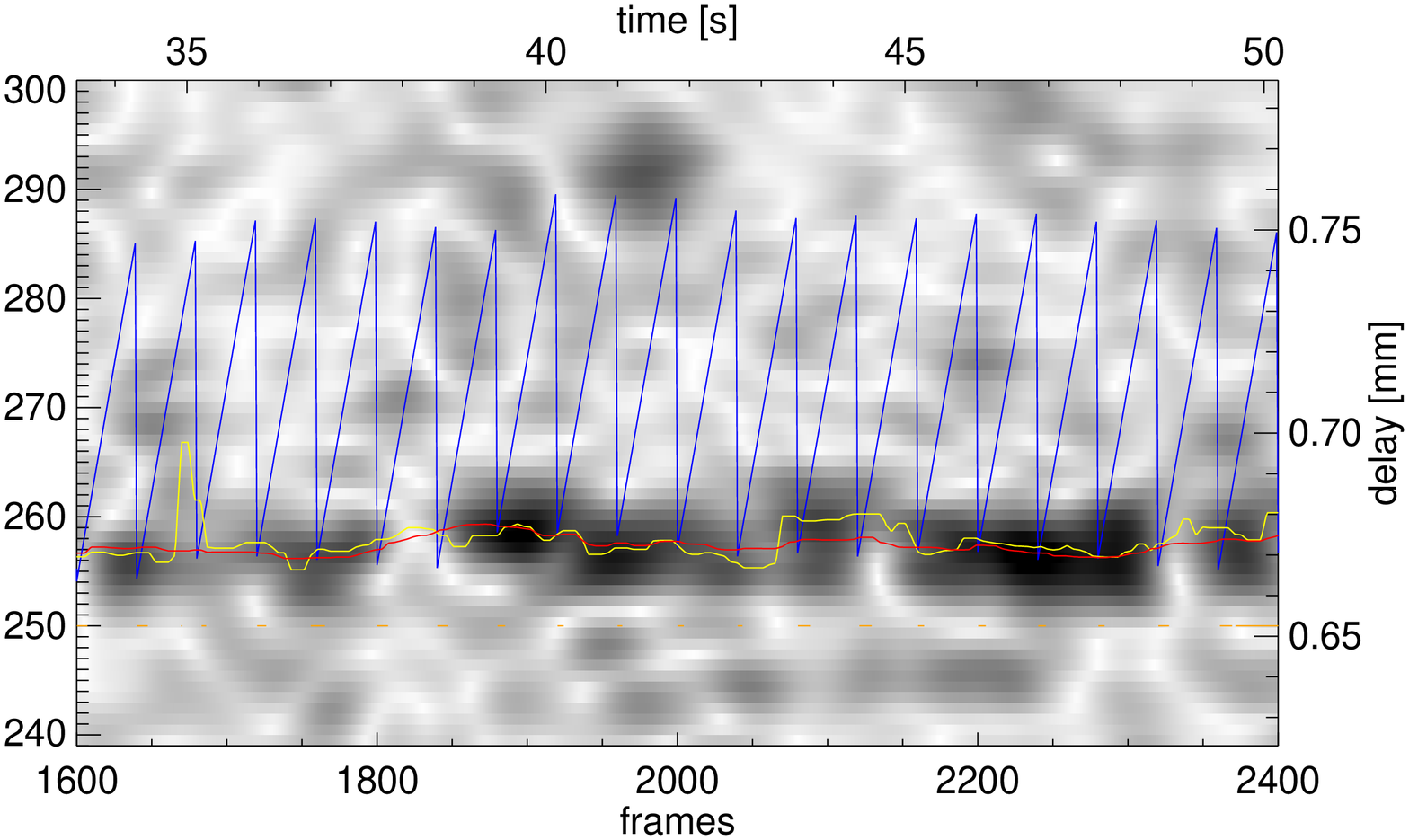}
\end{tabular}
\end{center}
\caption{ \label{fig:groupdelay} Estimation of group delay with midivispipe (top) vs. faintvispipe (bottom). The backdrop images are identical (but have different sampling) and are the smoothed Fourier transforms of each frame after derotation of the instrumental delay. They show data from a bright calibrator, diluted to represent a 150 mJy source (see Section \ref{sec:corrloss}). The red line is the ``true'' group delay as determined from the undiluted calibrator data; the yellow line is the group delay that the EWS routines midivispipe and faintvispipe found, respectively. Note that the old routines (top) produce an estimation of group delay for each frame whereas the new routines (bottom) only store an average value every few frames. The blue zig-zag pattern illustrates the modulation of the MIDI-internal OPD during the integration. Orange dots / lines are plotted when a frame / group of frames has been flagged as bad by {\tt oirAutoFlag}. The calibrator data is from 2011-04-20 / 03:04:36 and the noise was extracted above the spectrum of the weak science observation at 02:46:19 of the same night.}
\end{figure}

The data- and computation-heavy processing steps are implemented as C routines {\tt oir...} and the pipeline scripts and interactive tools are written in IDL. The new features in EWS 2.0 are available in the IDL procedure {\tt faintVisPipe}, the traditional routine remains available as {\tt midiVisPipe}. A comparison between the group delay determination of the two routines for simulated weak data (see Section \ref{sec:corrloss}) is shown in Fig.~\ref{fig:groupdelay}. The data reduction steps are:

\begin{enumerate}
	\item {\em Raw data} $\rightarrow$ {\tt oir1dCompressData} $\rightarrow$ {\em compressed.fits} \\ apply mask to raw data to reduce two-dimensional detector image to one-dimensional spectrum. {\bf New:} An optimized mask can be used to increase the signal/noise of the extracted spectrum. This mask is generated from the calibrator's ``fringe image'', i.e.\ a coherent image of the detector (for details see the IDL routine {\tt midiMakeMask}). This is particularly useful if no ``photometry'' observations have been made so that the default mask creation algorithm fails.
	\item {\em compressed.fits} $\rightarrow$ {\tt oirFormFringes} $\rightarrow$ {\em fringes.fits} \\ apply high-pass filter to remove non-modulated noise. For very low S/N sources, the data should be taken in ``offset tracking'' mode where the OPD does not cross ZOPD, but is modulated a few wavelengths offset from ZOPD. This has the advantage that the target spectrum becomes spectrally modulated. Thus, by subtracting the average from each frame, the background can be further suppressed.
	\item {\em fringes.fits} $\rightarrow$ {\tt oirRotateInsOpd} $\rightarrow$ {\em insopd.fits} \\ remove instrumental OPD variations (data now complex). The instrumental OPD variations are shown as a blue zig-zag line in Fig.~\ref{fig:groupdelay}.
	\item {\em insopd.fits} $\rightarrow$ {\tt oirFGroupDelay} $\rightarrow$ {\em groupdelay1.fits} \\ compute Fourier transform and searches maximum Fourier power in delay space per frame to determine the group delay. {\bf New:} Several minor and major changes. The most significant change to the pipeline is that this function can now be called twice (if the {\tt twopass} option is enabled). Most of the new features are only relevant in the second iteration (see below). The Fourier transformed raw data are shown as the backdrop image in Fig.~\ref{fig:groupdelay}.
	\item {\em groupdelay1.fits} $\rightarrow$ {\tt oirPowerDelay} $\rightarrow$ {\em powerdelay.fits} \\ {\bf New function:} make a crude estimate of the group delay for very low S/N data. Since the complex signal is uncorrelated over times larger than about the atmospheric coherence time ($\approx$ 20 ms at 10 $\um\!$), the estimate of the group delay is made incoherently on the amplitude of the complex data over a much longer time range (default 16 s). This crude estimate is used in the next step to constrain the range in which a finer group delay search is performed.
	\item {\em groupdelay1.fits, powerdelay.fits} $\rightarrow$ {\tt oirFGroupDelay} (2nd pass) $\rightarrow$ {\em groupdelay.fits} \\ {\bf New:} If called twice, this function searches for the maximum in the complex smoothed image in a region (default: $\pm$ 40 \um) around a crude estimate of the group delay determined by {\em powerdelay.fits}. This takes into account the empirical knowledge that the true position of ZOPD in the $N$ band typically does not vary more than $\sim$ 1 \um/s and has the advantage of avoiding noise peaks in very low S/N data.
	
	In order to increase the S/N for the group delay estimation, many frames (typically $\approx$ 20) have to be averaged together. To maximize the S/N, we now allow the group delay to vary within the smoothing interval. The found group delay is finally median smoothed, in order to suppress unphysical OPD jumps. Additionally, a previously existing bias in the group delay and phase estimation is reduced by excluding a certain time interval around the data currently being processed (see the EWS webpage for details). The found group delay is shown as the red (bright source) and yellow lines (faint source) in Fig.~\ref{fig:groupdelay}
	\item {\em fringes.fits, groupdelay.fits} $\rightarrow$ {\tt oirRotateGroupDelay} $\rightarrow$ {\em ungroupdelay.fits} \\ remove both the known instrumental delay and the group delay as determined above and estimate a phase offset induced by water vapor dispersion.
	\item {\em fringes.fits, groupdelay.fits} $\rightarrow$ {\tt oirAutoFlag} $\rightarrow$ {\em flag.fits} \\ determine the quality of each frame. The difference between the estimated OPD and the tracking delay of a good frame must not be larger than a certain value (default: 100 \um), but it must also not be smaller than a certain value (default: 10 \um), in case the source has been observed in ``off-zero tracking''. {\bf New:} The local jitter must not be larger than a certain value (default: 1.5 \um) to avoid significant correlation losses. Frames / groups of frames flagged as ``bad'' are shown with orange dots / lines in Fig.~\ref{fig:groupdelay}.
	\item {\em ungroupdelay.fits, flags.fits} $\rightarrow$ {\tt oirAverageVis} $\rightarrow$ {\em corr.fits} \\ coherently average all frames flagged as good.
\end{enumerate}

In essence, the new routines keep more frames for the averaging process at the end and remove biases in the data reduction process that become relevant for weak sources.

\subsection{Estimation of correlation losses}
\label{sec:corrloss}
For coherent integration, i.e.\ the data reduction method outlined above, a robust determination of the group delay for each frame is required since jittering leads to a reduction of the estimated correlated flux, so called correlation losses. A Gaussian error on the group delay of standard deviation $\sigma_{\rm GD}$ leads to a reduction of the correlated flux by the factor\cite{meisner2004} $\exp(-2 (\pi\sigma_{\rm GD}/\lambda)^2)$, e.g.\ $\sigma_{\rm GD} = 0.5 \um$ lead to 5~\% correlation loss and $\sigma_{\rm GD} = 1.6 \um$ lead to 40~\% correlation loss at $\lambda = 10 \um$.

\begin{figure}
\begin{center}
\begin{tabular}{c}
\includegraphics[width=6.75cm, angle=270]{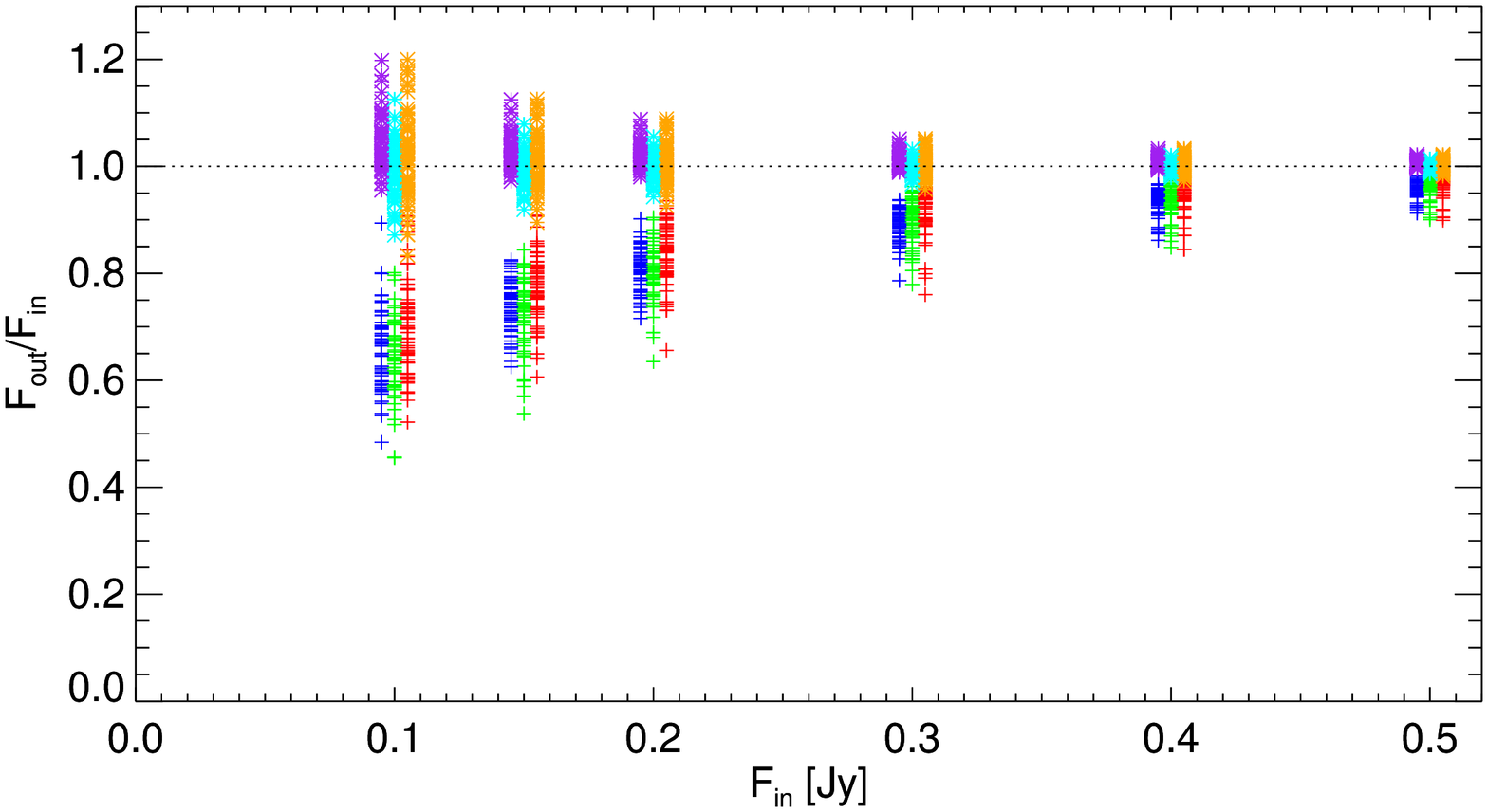}
\end{tabular}
\end{center}
\caption{ \label{fig:jitter} Recovered flux $F_{\rm out}$ as a fraction and function of input flux $F_{\rm in}$ for our dilution experiment and for the control experiment. The colors denote the wavelengths 8.5 \um (blue / purple), 10.5 \um (green / cyan) and 12.5 \um (red / orange) for the case with diluted and undiluted group delay estimation, respectively (see text); for each value of $F_{\rm in}$, the symbols for the three wavelengths have been shifted slightly for better readability. In total, 62 calibrators in three nights have been diluted to 100, 150, 200, 300, 400, 500 and 1000 mJy and the output flux has been determined as a function of input flux relative to the 1000 mJy experiment.}
\end{figure} 

Since, even for weak science targets, one typically uses bright ($F_{\nu} \gtrsim$ 5 Jy) calibrators to keep the errors in the calibration as small as possible, the group delay estimation for the target observation has a larger $\sigma_{\rm GD}$ than that of the calibrator observation and the correlation loss is not calibrated.

One way to address this problem is to observe calibrators of similar brightness as the science sources to study correlation losses. While this has been used successfully to show that correlation losses are below 5\% for stars with fluxes as low as $\approx$ 300 mJy\cite{kishimoto2011b}, correlation losses for the faintest science targets observable with the VLTI cannot be tested with this method since, for very faint stars, precise mid-IR flux estimates are not available. The 12 \um fluxes from the IRAS Faint Source Catalogue are typically uncertain by more than 10\% for fluxes lower than 200 mJy and AKARI/IRC Point Source Catalogue flux uncertainties at 9 \um are still considerably larger than 5 \% for fluxes as weak as 200 mJy. However, as is shown below, significant correlation losses only appear at such low fluxes.

A different approach to estimate correlation losses in the data reduction of faint targets is to use a bright target observation and ``dilute'' it. This is done by adding to a noise file the raw data of a target with known flux $F_{\nu, c}$ (i.e.\ a calibrator), multiplied by a factor $f \ll 1$ to simulate an observation of a known, weak input flux $F_{\rm in} = f \cdot F_{\nu, c}$. In this way not only the input flux is precisely known but also a very accurate estimation of the group delay from the reduction of the undiluted target. Of course this method cannot simulate the AO and lab guiding (IRIS) performance of a weak AGN observation, but this is also not possible by observing faint stars since stars have blue infrared colors, while AGNs are red. As the various VLTI subsystems (Adaptive Optics, lab guiding) operate on visual or near-infrared light, they work better for weak stars than for weak AGNs.

The result of such a simulated observation of a weak target is shown in Fig.~\ref{fig:groupdelay} where the backdrop image is from the observation of a bright calibrator star that has been mixed with noise in such a way to simulate a source with a flux of 150 mJy. It can be seen that the traditional routine (top image) has significant problems in finding the true group delay whereas the optimized routine has lower scatter so that less frames are flagged as bad. However, as can be seen in the lower part of Fig.~\ref{fig:groupdelay}, there are still significant deviations from the true group delay. They lead to reduced correlated fluxes compared to the known input flux for very weak sources as can be seen in Fig.~\ref{fig:jitter}. In this experiment, 62 calibrators of three nights have been diluted to represent weak targets with 100 - 1000 mJy of correlated flux and the correlation losses relative to the 1000 mJy case have been determined at three wavelengths (8.5, 10.5, 12.5 \um). It can be seen that the decorrelation is stronger for the blue than for the red part of the spectrum, as expected, but that the spread in decorrelation is larger at the long-wavelength end.

\begin{table}[t]
\vspace{1cm}
\begin{center}
\begin{tabular}{|c|ccc|ccc|}
$F_{\rm in}$ &\multicolumn{3}{c}{$F_{\rm out}/F_{\rm in}$ (faint data + faint GD)} & \multicolumn{3}{c}{$F_{\rm out}/F_{\rm in}$ (faint data + true GD)}\\
$[{\rm Jy}]$ & 8.5 \um & 10.5 \um & 12.5 \um & 8.5 \um & 10.5 \um & 12.5 \um\\
\hline
0.500&0.960$\pm$0.002&0.963$\pm$0.003&0.973$\pm$0.003&1.005$\pm$0.001&1.000$\pm$0.001&1.002$\pm$0.001\\
0.400&0.934$\pm$0.003&0.938$\pm$0.004&0.953$\pm$0.004&1.008$\pm$0.001&0.999$\pm$0.001&1.003$\pm$0.002\\
0.300&0.891$\pm$0.004&0.894$\pm$0.005&0.920$\pm$0.006&1.012$\pm$0.002&0.999$\pm$0.002&1.004$\pm$0.003\\
0.200&0.809$\pm$0.005&0.800$\pm$0.007&0.842$\pm$0.007&1.020$\pm$0.003&0.998$\pm$0.003&1.007$\pm$0.005\\
0.150&0.738$\pm$0.006&0.724$\pm$0.008&0.777$\pm$0.008&1.029$\pm$0.004&0.998$\pm$0.004&1.010$\pm$0.006\\
0.100&0.657$\pm$0.009&0.641$\pm$0.009&0.708$\pm$0.012&1.046$\pm$0.006&0.996$\pm$0.007&1.016$\pm$0.010\\
\end{tabular}
\end{center}
\caption{\label{tab:decorr}Correlation losses as a function of wavelength and input flux for the case with diluted and undiluted group delay estimation, respectively (see Section \ref{sec:corrloss}). The numbers quoted here depend on data reduction parameters. They have been computed using all 62 calibrator observations of 2010-03-26, 2010-05-27 and 2011-04-20 and with the EWS snapshot version 2012 Jan 25. The uncertainties are given as the error of the mean.}
\end{table}

We tested our dilution procedure by reducing a diluted input file with the group delay (and phase) estimation from the undiluted file and found no correlation losses (as expected). This means the reason for the losses is really the erroneous determination of group delay and phase in the weak sources and not caused by the dilution procedure itself. The correlation losses and the numbers for the control experiment are compiled in Table~\ref{tab:decorr}.

For the weakest sources, we therefore correct for the decorrelation losses by multiplying the fluxes that the data reduction routine finds with a decorrelation correction factor which we derived empirically from 62 observations of calibrators in three different nights (see Fig.~\ref{fig:jitter} and Table~\ref{tab:decorr}). Since the spread in $F_{\rm out}/F_{\rm in}$ is approximately the same for the dilution experiment with diluted and original group delay file (the blue/purple, green/cyan and red/orange pairs), we attribute this spread to the usual photon noise error that is already included in the error budget and derive the error for the decorrelation correction as the error of the mean of all realizations. This error is very small ($\lesssim 1 \%$ for all fluxes, see Table~\ref{tab:decorr}), it can be neglected for our error estimation (Eq. \ref{eq:errsum}).





\subsection{Data selection / quality control}

\begin{figure}
\begin{center}
\begin{tabular}{c}
\includegraphics[width=5cm, angle=270]{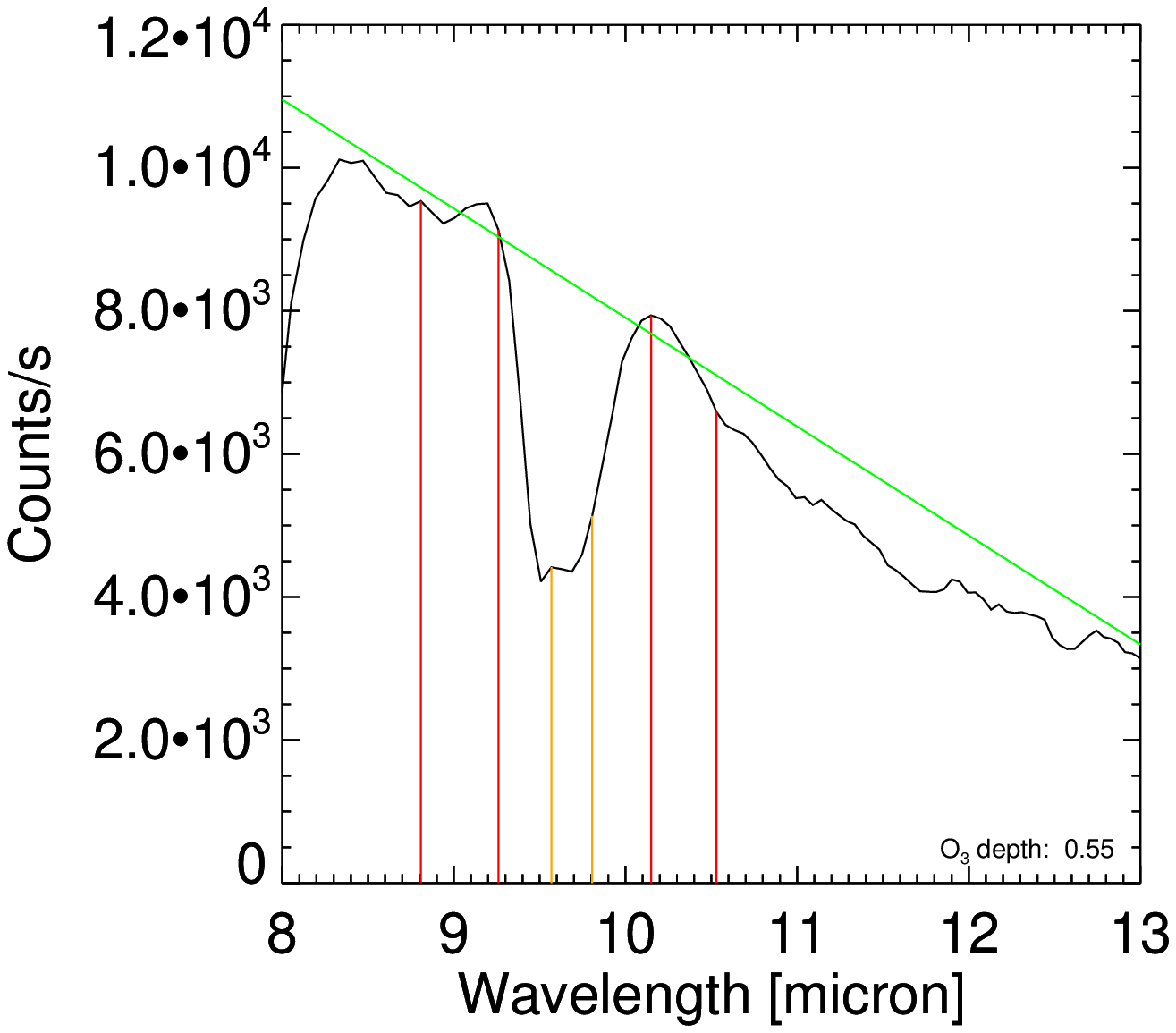}
\includegraphics[width=5cm, angle=270]{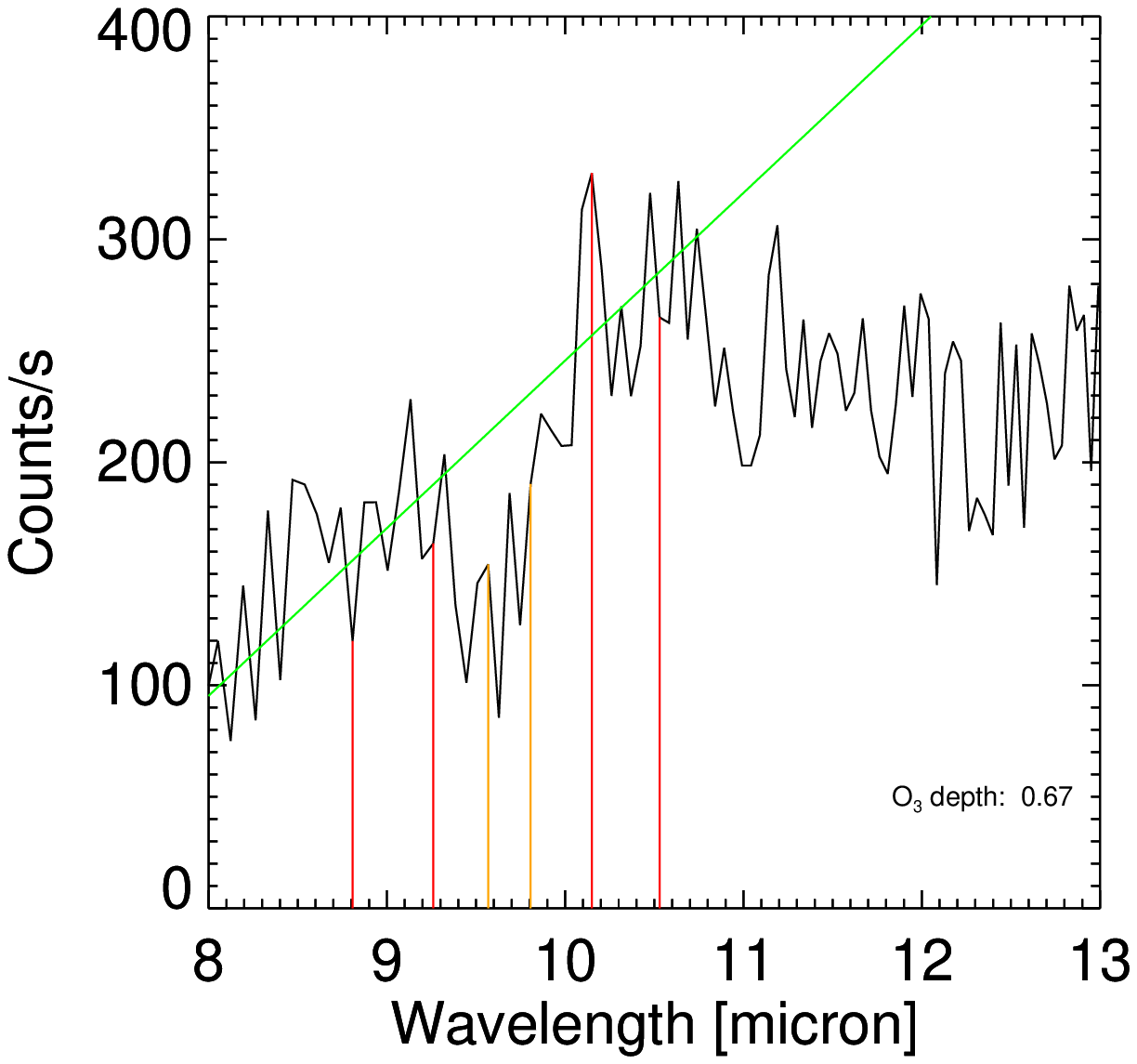}
\includegraphics[width=5cm, angle=270]{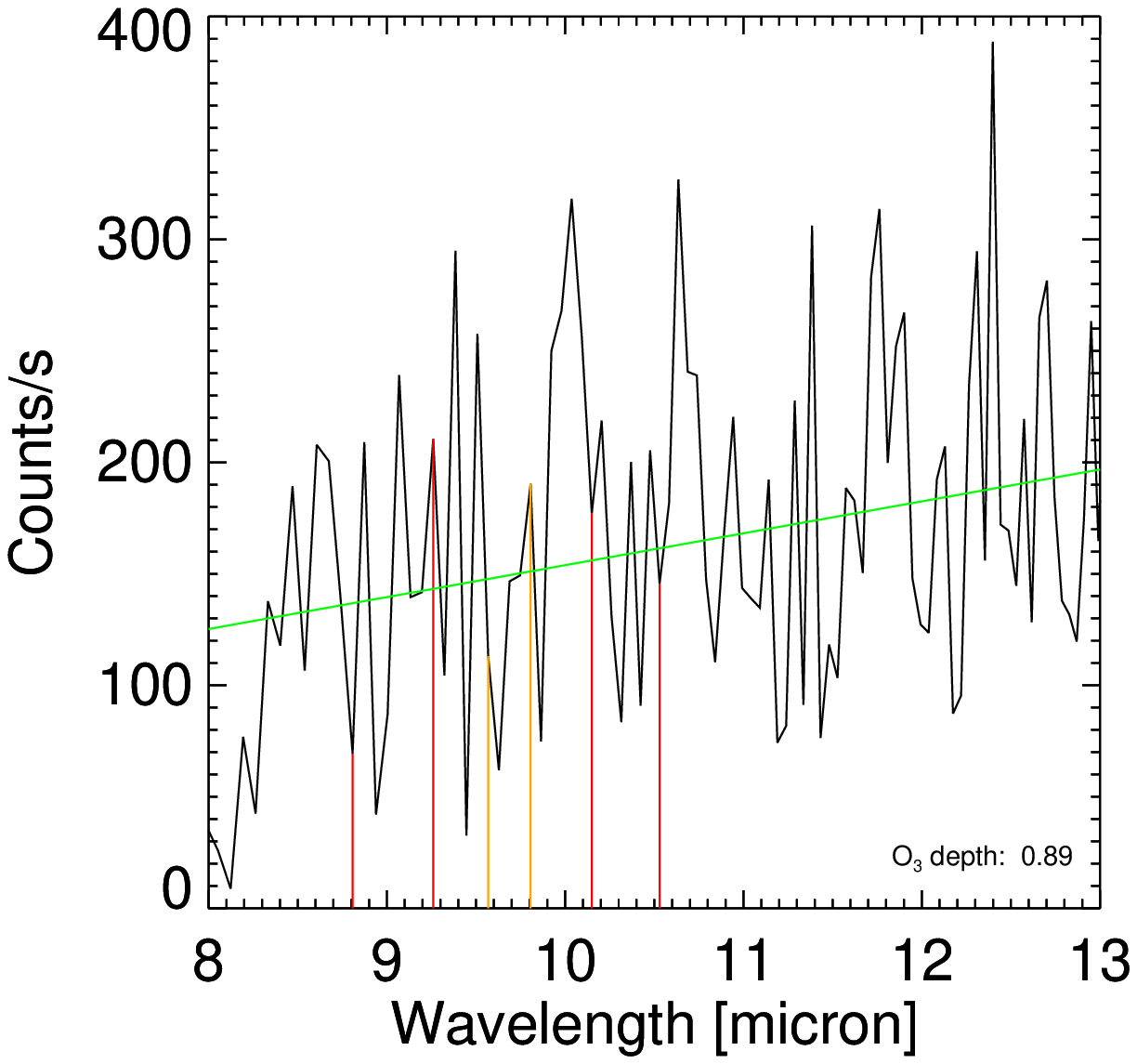}
\end{tabular}
\end{center}
\caption{ \label{fig:o3} Correlated flux count rate as a function of wavelength for three different targets of the night of 2011-08-17: a bright calibrator (06:26:11, left panel), a good observation of a very faint target (middle, NGC 7469 at 06:36:38, correlated flux at 12 \um $\approx$ 250 mJy) and a bad observation of the same target (right, 08:39:24). To determine the quality of the reduced data, the flux is interpolated between about 9.0 and 10.3 \um (interpolation regions marked with red lines) to estimate the ``continuum'' flux outside the atmospheric ozone absorption feature at 9.6 \um (orange). The linear interpolation is shown in green and the flux inside the feature as a fraction of the interpolated flux at this wavelength is printed in the lower right corner of each panel. Good observations typically have values smaller than 0.76}
\end{figure}

Apart from the frame selection in EWS, which automatically de-selects frames for the averaging process if the group delay estimation is likely to be wrong, we also introduce a track selection using the atmospheric ozone absorption feature at 9.6 \um. Any radiation from outside the earth's atmosphere must have this feature imprinted on its mid-IR spectrum leading to a reduction in flux at approximately this wavelength. We estimate the continuum flux at this wavelength by interpolating between two wavelengths regions that are unaffected by atmospheric ozone absorption and determine the count rate in the feature as a fraction of the interpolated count rate (see Fig.~\ref{fig:o3}). If this fraction is {\em larger} than a threshold value, the flux is not dominated by emission from outside the atmosphere and the observation is flagged as ``bad''. We found that a threshold value of 0.76 gives results that are most consistent with other estimators of data quality.

We tested to additionally select data by seeing and airmass, however the simple ``ozone test'' was sufficient to de-select most bad fringe tracks. Additionally a few tracks were de-selected manually where clouds might have affected the flux, seeing was excessively large ($>$ 2\arcsec) or the signal/noise was extremely low.

\section{Error estimation}

For bright sources, the uncertainty in the correlated flux can be estimated by dividing the frames to be averaged for the estimation of the correlated flux in a number of bins (default: 16) and by determining the rms of these. This is a good approximation of the photon noise error. However, it does not take into account the uncertainty in the conversion factor in the case of direct flux calibration and it is not corrected for possible correlation losses.

\subsection{Conversion factor uncertainty}

\begin{figure}
\begin{center}
\begin{tabular}{c}
\includegraphics[width=5.85cm, angle=270]{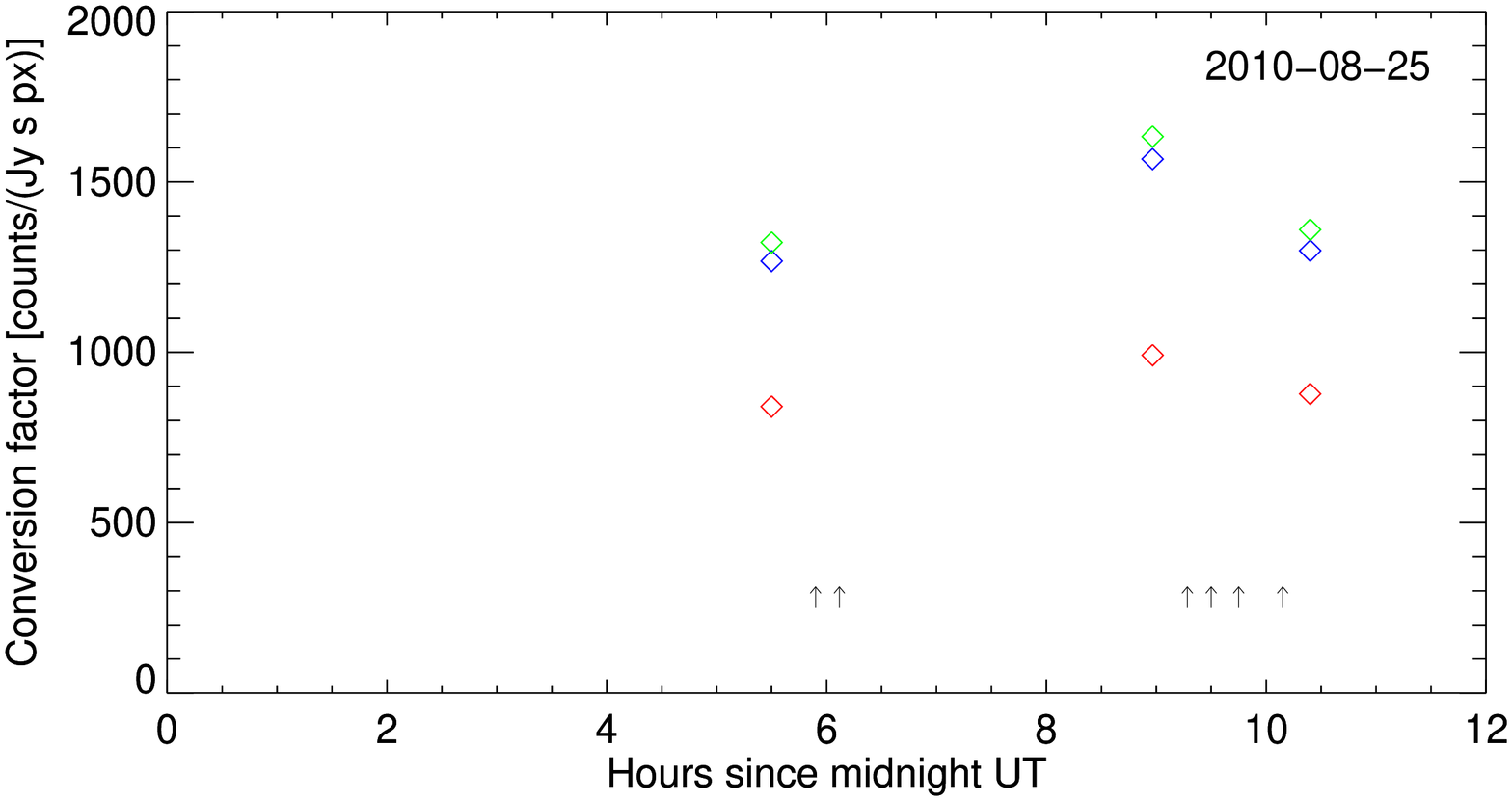}\\
\includegraphics[width=5.85cm, angle=270]{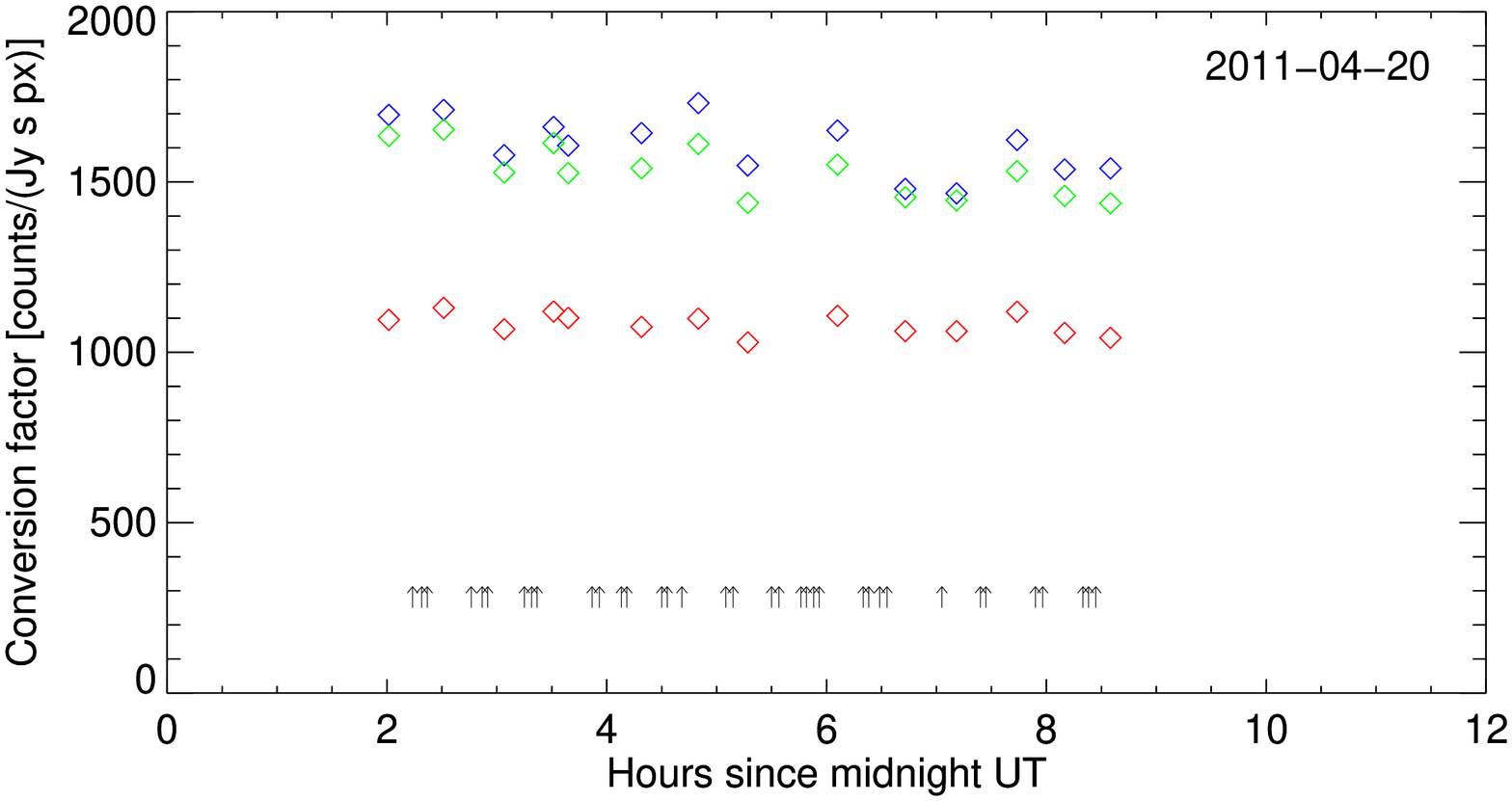}\\
\includegraphics[width=5.85cm, angle=270]{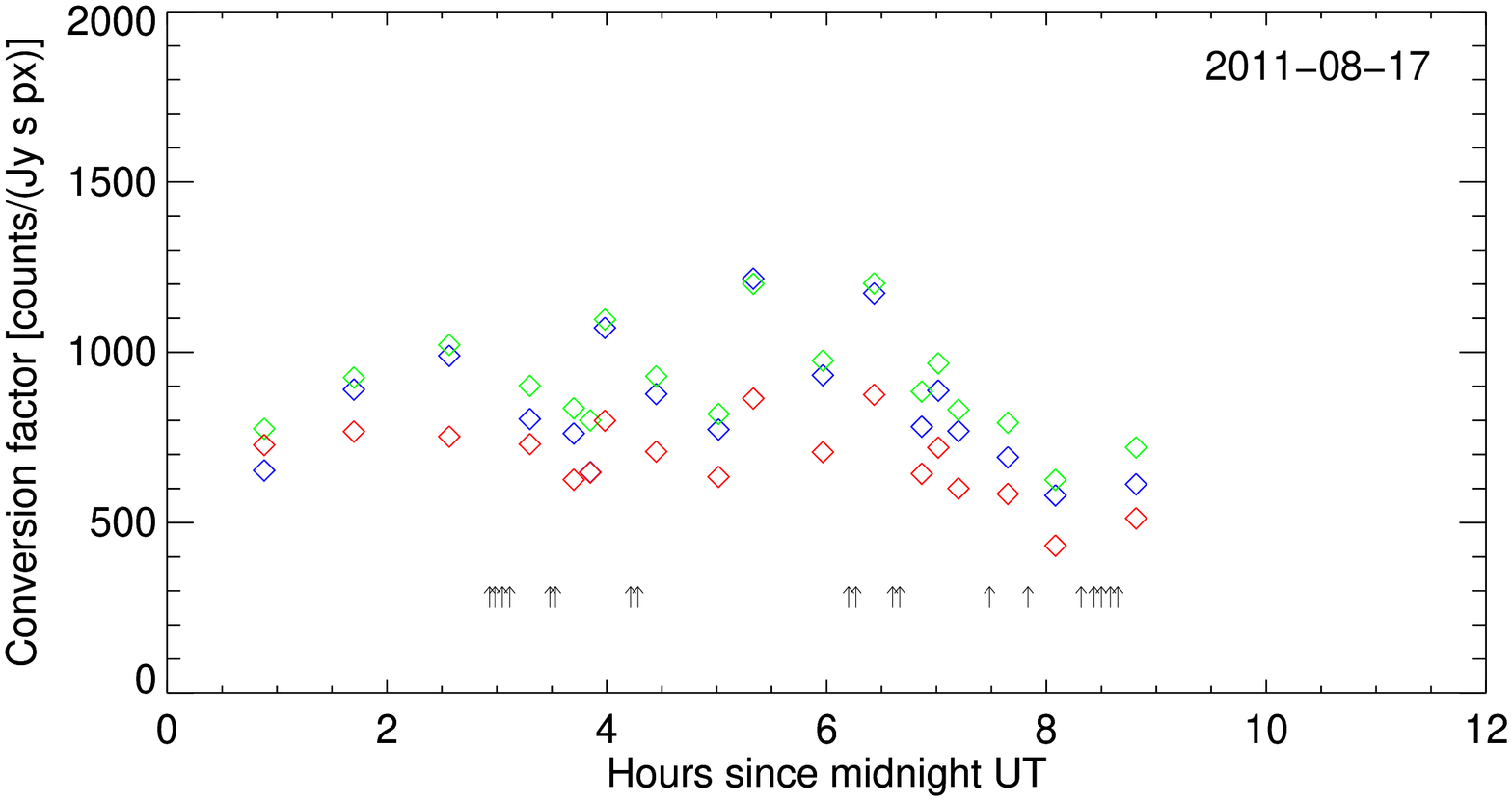}
\end{tabular}
\end{center}
\caption{ \label{fig:gain} Correlated flux conversion factor (counts/(Jy s px)) as a function of time for three nights: 2010-08-25 (top), 2011-04-20 (middle), 2011-08-17 (bottom). The colored diamonds show the conversion factor at 8.5 (blue), 10.5 (green) and 12.5 \um (red), the small black arrows indicate the times at which science observations have been taken. Only with a dense sampling of the correlated flux can the stability of the conversion factor be estimated reliably. In 2010-08-25 the (airmass-corrected) standard deviation at 12.5 micron was 8.1 \%, but is badly sampled and therefore a bad estimate for the calibration uncertainty due to the direct flux calibration. In 2011-04-20 it was 2.2 \% and in 2011-08-17 it was 16.1 \%. Since these nights have been densely sampled, these numbers are more trustworthy to use as calibration uncertainty.\newline
In the night of 2011-04-20, where only correlated fluxes have been observed (no acquisition, almost no fringe searches and no photometries), a total of 14 calibrator observations and 36 science observations have been taken in a little less than 6.5 hours, leading to the above quoted number of 8 minutes per observation.}
\end{figure} 

\begin{figure}
\begin{center}
\begin{tabular}{c}
\includegraphics[width=5cm, angle=270]{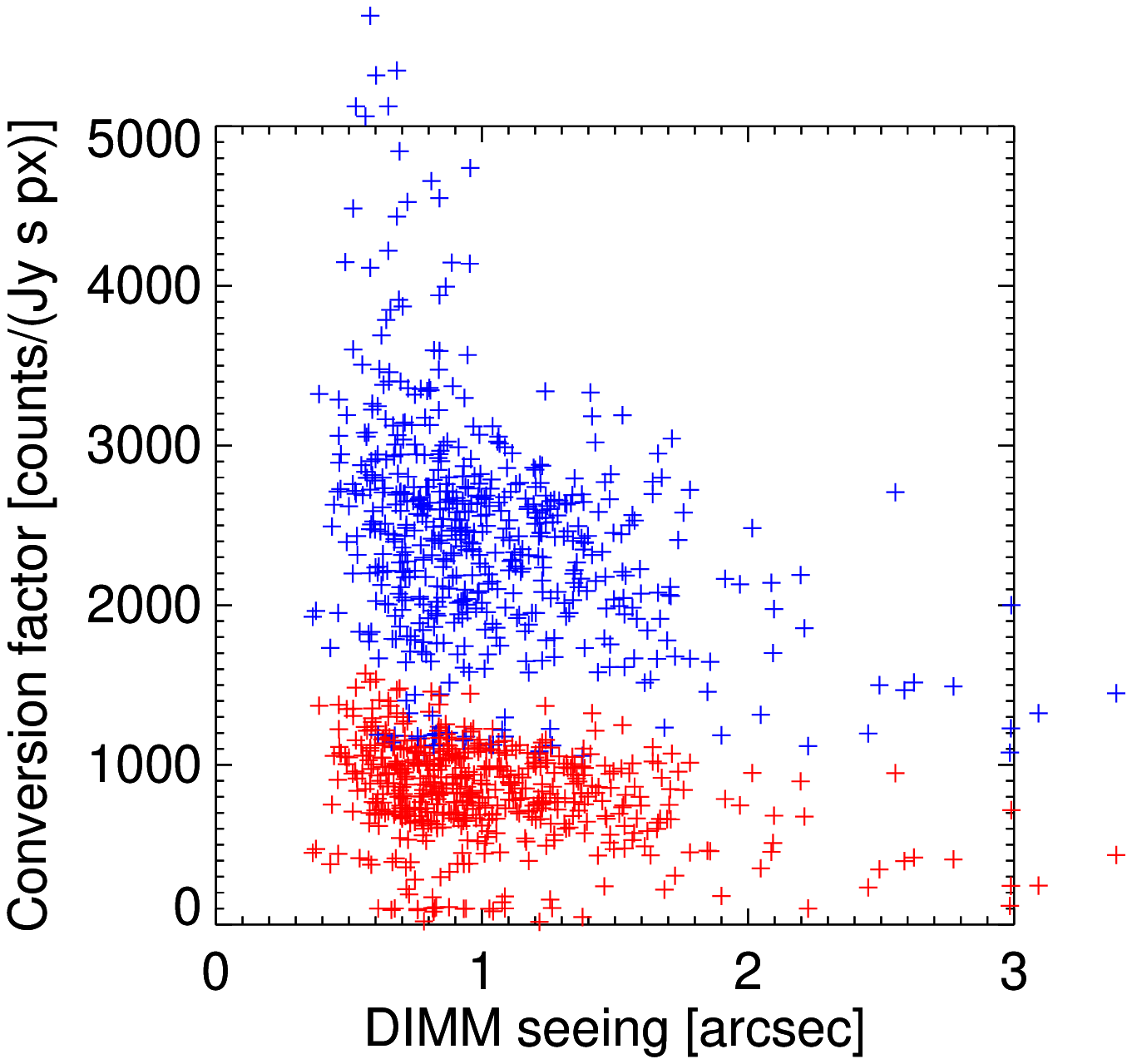}
\includegraphics[width=5cm, angle=270]{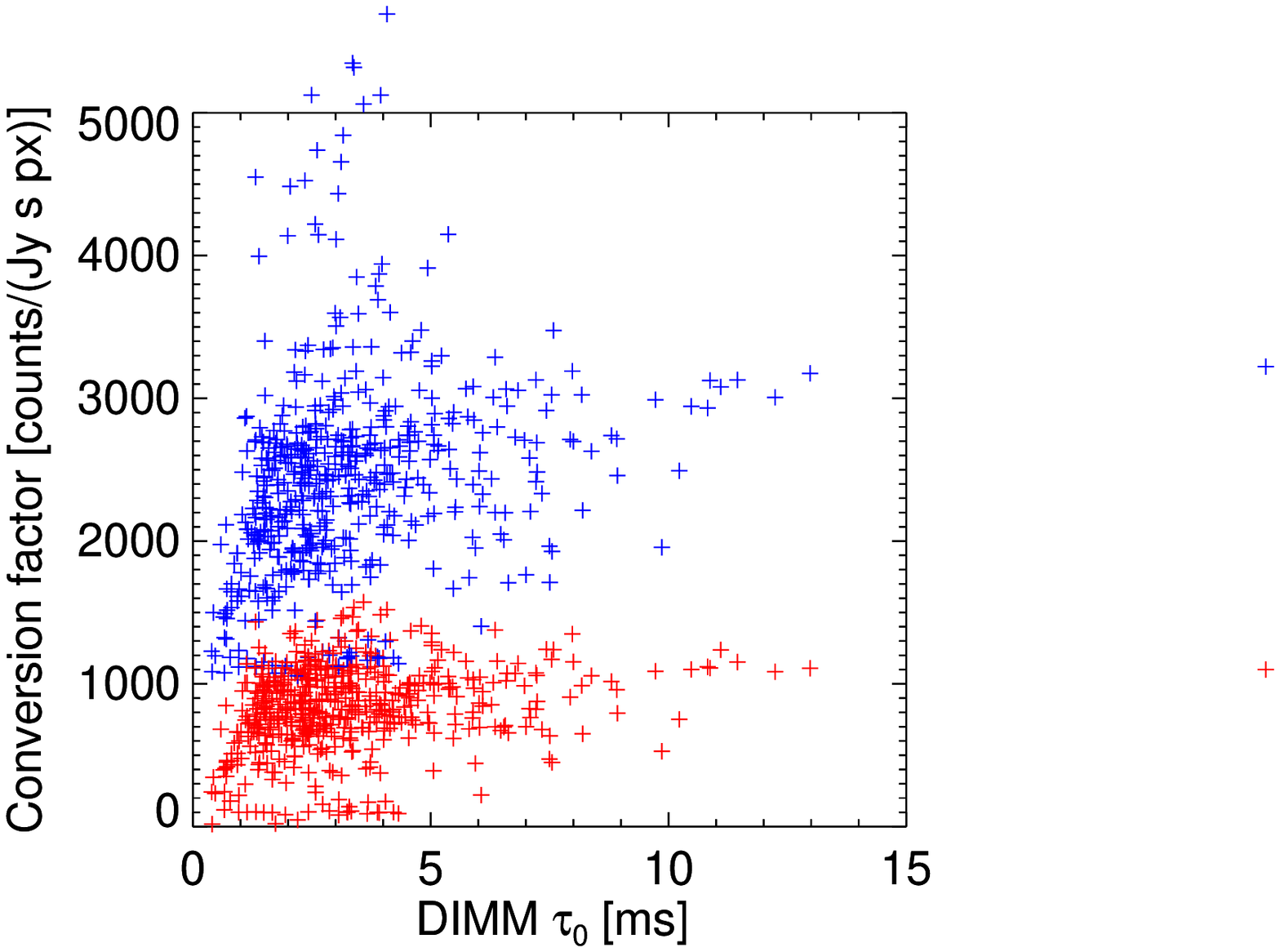}
\includegraphics[width=5cm, angle=270]{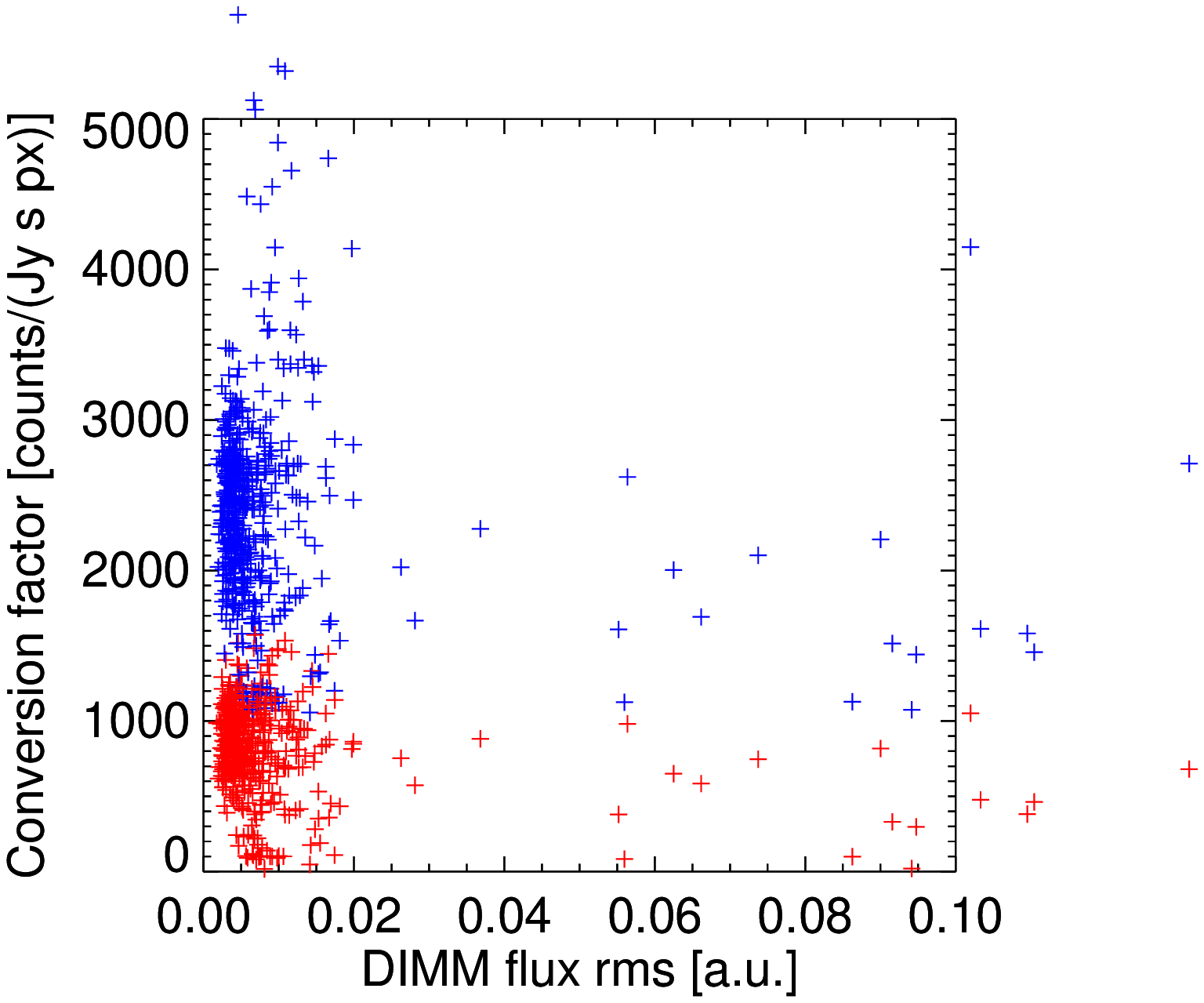}
\end{tabular}
\end{center}
\caption{ \label{fig:gaincorr} Correlated flux conversion factor (counts/(Jy s px)) as a function of DIMM seeing (left), DIMM $\tau_0$ (center) and DIMM flux rms (right), all measured in the visual. The latter can be considered as a proxy for the cloud coverage. Red and blue crosses denote the conversion factor values at 12.5 and 8.5 \um, respectively. They are offset by 1000 counts/(Jy s px) for clarity. There are weak correlations in the expected sense (higher conversion factors for better seeing, larger $\tau_0$, lower cloud coverage), but the correlations are weak and there is large scatter. The data show more than 600 calibrator measurements from 2005 -- 2011.}
\end{figure} 

For faint sources, the ``direct flux calibration'' method is preferable compared to the visibility calibration since it avoids the noisy measurement of the total flux (see Section \ref{sec:obs}). However, the flux calibration method is sensitive to variations in the atmospheric and VLTI transfer function, i.e.\ the conversion factor. Using the standard observing scheme, it is only measured every hour which leaves a significant uncertainty about the behavior of the conversion factor between the calibrator and the science observation. When omitting all single-dish observations and only observing fringes, the conversion factor can be sampled much more densely and a more accurate estimate of its variance can be obtained (see Fig.~\ref{fig:gain}). We use the conversion factor of the calibrator closest to the science object in time, airmass and distance on sky and estimate its error, $\sigma_{\rm CF}$, as the standard deviation of the conversion factor derived from all measurements over the night. Variations of the conversion factor with airmass contribute little to its total scatter. We nevertheless remove them by fitting a line to the $\log$ CF -- airmass relation and subtracting the value of this function for each measurement before calculating the standard deviation.

During the course of the MIDI AGN Large Programme, the relative error of the conversion factor was as low as $\approx$ 3 \% for excellent nights and as large as $\approx$ 30\% in bad nights. The median airmass-corrected value of the relative conversion factor error was 8.9 \%. It is typically larger at the red end of the atmospheric $N$ band. For most observations, this is the dominant source of error and we therefore investigated correlations of the conversion factor with other observables that are sampled more frequently by the DIMM in the visual, such as the seeing, the atmospheric coherence time $\tau_0$ and the sky transmission (Fig.~\ref{fig:gaincorr}). It can be seen that correlations in the expected sense do exist (lower conversion factors for larger seeing, smaller $\tau_0$ and lower sky transmission), but that the correlations are weak and have very strong scatter making them unusable to reduce the conversion factor error. Possible other correlations include the precipitable water vapor (PWV) above the observatory. We investigated this and found another weak correlation in the expected sense, but since the PWV was (until recently) only monitored once per night or more rarely, it is hard to constrain this correlation, if it exists.

\subsection{Error estimation for averaged fluxes}

\begin{figure}
\begin{center}
\begin{tabular}{c}
\includegraphics[width=10cm, angle=270]{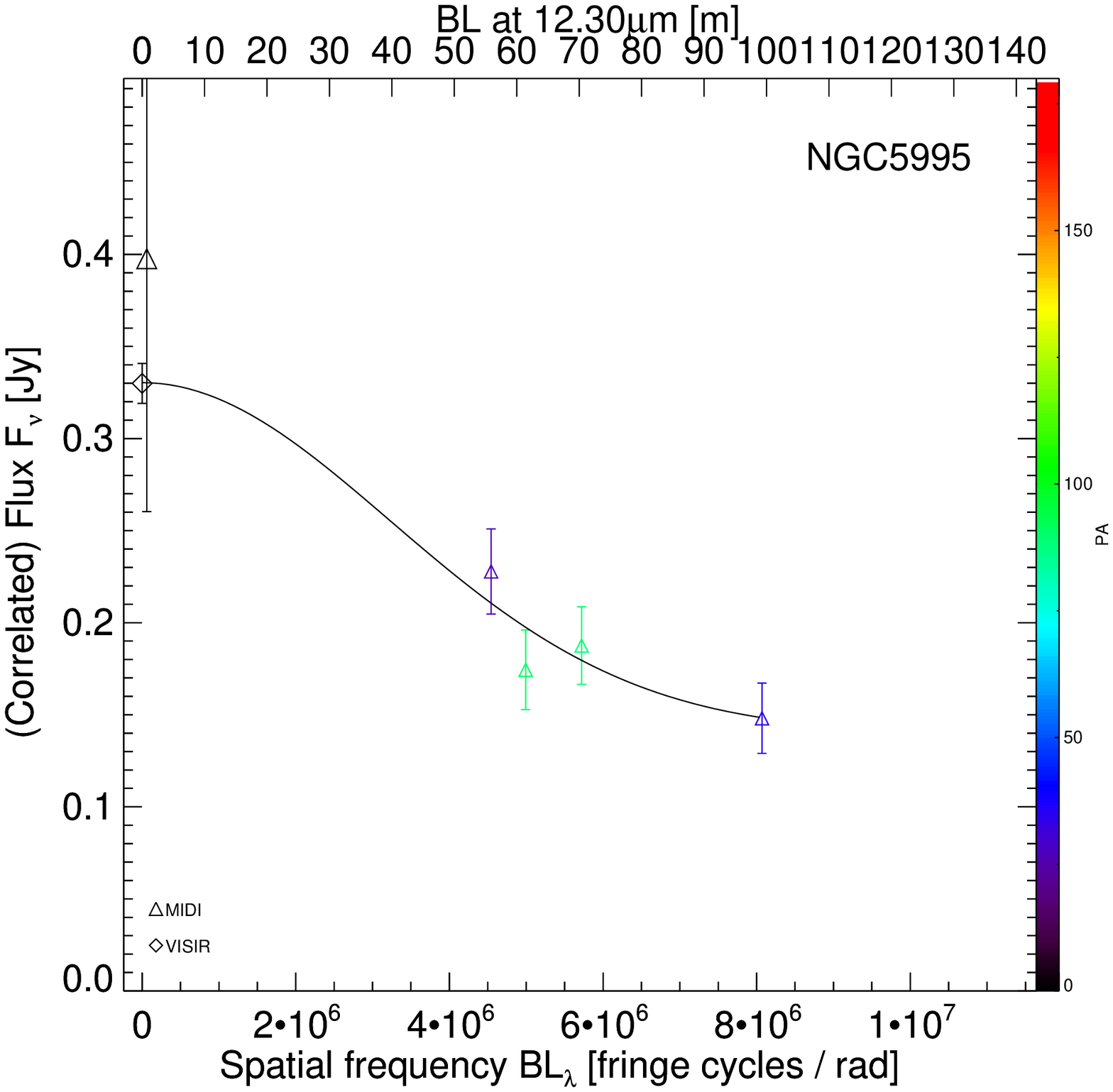}
\end{tabular}
\end{center}
\caption{ \label{fig:radial} Correlated flux as a function of projected baseline length for the Seyfert 2 galaxy NGC 5995. The two points at zero baseline are total flux measurements from MIDI (triangle, large error bar) and VISIR (diamond). The four colored points represent correlated flux observations at $\lambda = $12.3 \um restframe at different position angles (see color bar at the right) in the nights of 2010-03-26 (Baseline: 55m), 2010-03-27 (100 m) and 2011-04-17 (60m and 70m). The error bars of the correlated flux measurements take into account the three noise and uncertainty sources: photon noise, conversion factor variance and the uncertainty in the correction of correlation losses. They demonstrate that consistent results can be achieved with MIDI using our data reduction and calibration methods for fluxes as low as $\approx$ 150 mJy.}
\end{figure} 

For very weak targets, often only fluxes at a few points in the atmospheric $N$ band are given, e.g.\ at 8.5 $\pm$ 0.2 $\um\!$ and at 12.5 $\pm$ 0.2 $\um\!$. Our investigations have shown that, in addition to the photon noise error that is returned by EWS, $\sigma_{\rm phot}$,  the calibration uncertainty is a major source of error and has to be taken into account. We estimate the calibration uncertainty as the (airmass-corrected) standard deviation of the conversion factor per night (and per baseline) $\sigma_{\rm CF}$. The error of the averaged flux is then

\begin{equation}
	\label{eq:errsum}
	\sigma^2 = 1/N \cdot \sigma_{\rm phot}^2 + \sigma_{\rm CF}^2
\end{equation}

where $N$ is the number of wavelength channels that have been averaged.

This treatment of errors gives consistent results for sources as weak as $\approx$ 150 mJy, e.g.\ for the MIDI AGN Large Programme source NGC~5995 (Fig.~\ref{fig:radial}). For this source we fitted the data to a simple one-dimensional visibility model of an unresolved point source and a resolved source of a Gaussian intensity profile (similar to a one-component power-law intensity profile). The spread of the data about this model is compatible with the error bars derived using the error calculation presented above, i.e.\ the $\chi^2$ is approximately equal to the number of degrees of freedom.

 





\section{Summary and applicability to other targets}

The weakest objects that are observable with MIDI at the VLTI require special observing and data reduction methods in order to limit and reliably estimate the otherwise large errors. The largest improvement is achieved by calibrating correlated fluxes directly (instead of visibilities). In order to limit the uncertainty due to the sometimes very variable transfer function, the target and calibrator observations should be taken as close in time as possible which is achievable by concentrating on only taking correlated fluxes. In the data reduction, the biggest challenge for weak sources is an accurate estimation of the group delay. Progress has been made by implementing an iterative approach for the group delay estimation and especially by simulating OPD changes during each smoothing interval.

A method has been developed to robustly test the data reduction method for weak sources by generating artificially diluted data of a real observation and comparing its reduction with the one using the original data. This way, significant correlation losses have been uncovered for fluxes lower than 300 mJy. They can be corrected for by applying a decorrelation correction factor that depends both on the observing conditions and the data reduction parameters.

After both photon noise and conversion factor errors have been taken into account and after correction of correlation losses, consistent results are achieved for weak targets. Using the methods described here we can reproducibly observe objects as faint as $\approx$ 150 mJy with an uncertainty of $\approx$ 15 \% under average conditions.

Some of the methods presented here are a part of EWS and are available from the previously quoted web page; the higher level methods, that require a knowledge of all observations of a night, such as the determination of conversion factor variations and decorrelation correction factors have not (yet) been implemented in EWS. They are available as a Google code project and can be downloaded from \href{http://code.google.com/p/miditools/}{http://code.google.com/p/miditools/}. Interaction with the authors is highly encouraged if you plan to use these scripts.

The observing and data reduction method as well as the error estimation are not specific to observations of weak AGNs but may be applied for MIDI observations of any weak target. With PRIMA becoming available as an external fringe tracker for MIDI observations\cite{mueller2010}, weak targets will be observed more frequently with MIDI and a precise estimation of the measurement uncertainties is needed.


\begin{thebibliography}{10}

\bibitem{antonucci1993}
{Antonucci}, R., ``{Unified models for active galactic nuclei and quasars},''
  {\em \araa}~{\bf 31},  473--521 (1993).

\bibitem{hoenig2010}
{H{\"o}nig}, S.~F., {Kishimoto}, M., {Gandhi}, P., {Smette}, A., {Asmus}, D.,
  {Duschl}, W., {Polletta}, M., and {Weigelt}, G., ``{The dusty heart of nearby
  active galaxies. I. High-spatial resolution mid-IR spectro-photometry of
  Seyfert galaxies},'' {\em \aap}~{\bf 515},  A23+ (June 2010).

\bibitem{ramosalmeida2011}
{Ramos Almeida}, C., {Levenson}, N.~A., {Alonso-Herrero}, A., {Asensio Ramos},
  A., {Rodr{\'{\i}}guez Espinosa}, J.~M., {P{\'e}rez Garc{\'{\i}}a}, A.~M.,
  {Packham}, C., {Mason}, R., {Radomski}, J.~T., and {D{\'{\i}}az-Santos}, T.,
  ``{Testing the Unification Model for Active Galactic Nuclei in the Infrared:
  Are the Obscuring Tori of Type 1 and 2 Seyferts Different?},'' {\em
  \apj}~{\bf 731},  92--+ (Apr. 2011).

\bibitem{nenkova2002}
{Nenkova}, M., {Ivezi{\'c}}, {\v Z}., and {Elitzur}, M., ``{Dust Emission from
  Active Galactic Nuclei},'' {\em \apjl}~{\bf 570},  L9--L12 (May 2002).

\bibitem{hoenig2006}
{H{\"o}nig}, S.~F., {Beckert}, T., {Ohnaka}, K., and {Weigelt}, G.,
  ``{Radiative transfer modeling of three-dimensional clumpy AGN tori and its
  application to NGC 1068},'' {\em \aap}~{\bf 452},  459--471 (June 2006).

\bibitem{nenkova2008}
{Nenkova}, M., {Sirocky}, M.~M., {Ivezi{\'c}}, {\v Z}., and {Elitzur}, M.,
  ``{AGN Dusty Tori. I. Handling of Clumpy Media},'' {\em \apj}~{\bf 685},
  147--159 (Sept. 2008).

\bibitem{schartmann2008}
{Schartmann}, M., {Meisenheimer}, K., {Camenzind}, M., {Wolf}, S., {Tristram},
  K.~R.~W., and {Henning}, T., ``{Three-dimensional radiative transfer models
  of clumpy tori in Seyfert galaxies},'' {\em \aap}~{\bf 482},  67--80 (Apr.
  2008).

\bibitem{jaffe2004}
{Jaffe}, W., {Meisenheimer}, K., {R{\"o}ttgering}, H.~J.~A., {Leinert}, C.,
  {Richichi}, A., {Chesneau}, O., {Fraix-Burnet}, D., {Glazenborg-Kluttig}, A.,
  {Granato}, G.-L., {Graser}, U., {Heijligers}, B., {K{\"o}hler}, R., {Malbet},
  F., {Miley}, G.~K., {Paresce}, F., {Pel}, J.-W., {Perrin}, G., {Przygodda},
  F., {Schoeller}, M., {Sol}, H., {Waters}, L.~B.~F.~M., {Weigelt}, G.,
  {Woillez}, J., and {de Zeeuw}, P.~T., ``{The central dusty torus in the
  active nucleus of NGC 1068},'' {\em \nat}~{\bf 429},  47--49 (May 2004).

\bibitem{tristram2007b}
{Tristram}, K.~R.~W., {Meisenheimer}, K., {Jaffe}, W., {Schartmann}, M., {Rix},
  H.-W., {Leinert}, C., {Morel}, S., {Wittkowski}, M., {R{\"o}ttgering}, H.,
  {Perrin}, G., {Lopez}, B., {Raban}, D., {Cotton}, W.~D., {Graser}, U.,
  {Paresce}, F., and {Henning}, T., ``{Resolving the complex structure of the
  dust torus in the active nucleus of the Circinus galaxy},'' {\em \aap}~{\bf
  474},  837--850 (Nov. 2007).

\bibitem{raban2009}
{Raban}, D., {Jaffe}, W., {R{\"o}ttgering}, H., {Meisenheimer}, K., and
  {Tristram}, K.~R.~W., ``{Resolving the obscuring torus in NGC 1068 with the
  power of infrared interferometry: revealing the inner funnel of dust},'' {\em
  \mnras}~{\bf 394},  1325--1337 (Apr. 2009).

\bibitem{burtscher2009}
{Burtscher}, L., {Jaffe}, W., {Raban}, D., {Meisenheimer}, K., {Tristram},
  K.~R.~W., and {R{\"o}ttgering}, H., ``{Dust emission from a parsec-scale
  structure in the Seyfert 1 nucleus of NGC 4151},'' {\em \apjl}~{\bf 705},
  L53--L57 (Sept. 2009).

\bibitem{tristram2012}
{Tristram}, K.~R.~W., {Schartmann}, M., {Burtscher}, L., {Meisenheimer}, K.,
  {Jaffe}, W., {Kishimoto}, M., {H{\"o}nig}, S.~F., and {Weigelt}, G., ``{The
  complexity of parsec-scaled dusty tori in AGN},'' {\em ArXiv e-prints}  (Apr.
  2012).

\bibitem{tristram2009}
{Tristram}, K.~R.~W., {Raban}, D., {Meisenheimer}, K., {Jaffe}, W.,
  {R{\"o}ttgering}, H., {Burtscher}, L., {Cotton}, W.~D., {Graser}, U.,
  {Henning}, T., {Leinert}, C., {Lopez}, B., {Morel}, S., {Perrin}, G., and
  {Wittkowski}, M., ``{Parsec-scale dust distributions in Seyfert galaxies.
  Results of the MIDI AGN snapshot survey},'' {\em \aap}~{\bf 502},  67--84
  (July 2009).

\bibitem{kishimoto2011b}
{Kishimoto}, M., {H{\"o}nig}, S.~F., {Antonucci}, R., {Millour}, F.,
  {Tristram}, K.~R.~W., and {Weigelt}, G., ``{Mapping the radial structure of
  AGN tori},'' {\em \aap}~{\bf 536},  A78 (Dec. 2011).

\bibitem{burtscher2011}
{Burtscher}, L., {\em {Mid-infrared interferometry of AGN cores}}, PhD thesis,
  {Naturwissenschaftlich-Mathematische Gesamtfakult\"at der
  Ruprecht-Karls-Universit\"at Heidelberg} (Mai 2011).

\bibitem{dewit2012}
De~Wit, W.~J., ``Midi user manual,'' Doc. No. VLT-MAN-ESO-15820-3519, Issue 90
  (03 2012).

\bibitem{meisner2004}
{Meisner}, J.~A., {Tubbs}, R.~N., and {Jaffe}, W.~J., ``{Coherent integration
  of complex fringe visibility employing dispersion tracking},'' in [{\em
  Society of Photo-Optical Instrumentation Engineers (SPIE) Conference
  Series}{\nolinebreak\hspace{0.1em}]},  {W.~A.~Traub}, ed., {\em Society of
  Photo-Optical Instrumentation Engineers (SPIE) Conference Series} {\bf 5491},
   725--+ (Oct. 2004).

\bibitem{jaffe2004b}
{Jaffe}, W.~J., ``{Coherent fringe tracking and visibility estimation for
  MIDI},'' in [{\em Society of Photo-Optical Instrumentation Engineers (SPIE)
  Conference Series}{\nolinebreak\hspace{0.1em}]},  {Traub}, W.~A., ed., {\em
  Society of Photo-Optical Instrumentation Engineers (SPIE) Conference Series}
  {\bf 5491},  715--+ (Oct. 2004).

\bibitem{tristram2007}
{Tristram}, K.~R.~W., {\em {Mid-infrared interferometry of nearby Active
  Galactic Nuclei}}, PhD thesis, Max-Planck-Institut f{\"u}r Astronomie,
  K{\"o}nigstuhl 17, 69117 Heidelberg, Germany (July 2007).

\bibitem{mueller2010}
{M{\"u}ller}, A., {Pott}, J., {Morel}, S., {Abuter}, R., {van Belle}, G., {van
  Boekel}, R., {Burtscher}, L., {Delplancke}, F., {Henning}, T., {Jaffe}, W.,
  {Leinert}, C., {Lopez}, B., {Matter}, A., {Meisenheimer}, K., {Schmid}, C.,
  {Tristram}, K., and {Verhoeff}, A.~P., ``{First results using PRIMA FSU as a
  fringe tracker for MIDI},'' in [{\em Society of Photo-Optical Instrumentation
  Engineers (SPIE) Conference Series}{\nolinebreak\hspace{0.1em}]},  {\em
  Society of Photo-Optical Instrumentation Engineers (SPIE) Conference Series}
  {\bf 7734} (July 2010).

\end{thebibliography}

\bibliographystyle{spiebib}   

\end{document}